\documentclass[prl,twocolumn,showpacs,superscriptaddress,floatfix]{revtex4}

\usepackage{ifthen}
\usepackage{lineno}
\usepackage{slashed}
\usepackage{epsfig}
\usepackage{amsmath}
\usepackage{comment}
\usepackage{longtable}
\usepackage{subfigure}
\usepackage{color}
\usepackage{graphicx}

\newcommand{\qqb}{\ensuremath{q\bar{q}}}

\newcommand{\ttb}{\ensuremath{t\bar{t}}}
\newcommand{\mttb}{\ensuremath{M_{t\bar{t}}}}
\newcommand{\ppb}{\ensuremath{p\bar{p}}}

\newcommand{\ba}{\begin{array}}
\newcommand{\ea}{\end{array}}
\newcommand{\bc}{\begin{center}}
\newcommand{\ec}{\end{center}}
\newcommand{\bn}{\begin{enumerate}}
\newcommand{\en}{\end{enumerate}}
\newcommand{\bq}{\begin{equation}}
\newcommand{\eq}{\end{equation}}
\newcommand{\bi}{\begin{itemize}}
\newcommand{\ei}{\end{itemize}}   
\newcommand{\bh}{\begin{math}}
\newcommand{\eh}{\end{math}}
\newcommand{\br}{\begin{flushright}}
\newcommand{\er}{\end{flushright}}
\newcommand{\bl}{\begin{flushleft}}
\newcommand{\el}{\end{flushleft}}
\newcommand{\bt}{\begin{tabular}}
\newcommand{\et}{\end{tabular}}
\newcommand{\flame}

\newcommand{\MET}{\mbox{$\raisebox{.3ex}{$\not\!$}E_T$}}
\newcommand{\METVEC}{\mbox{$\raisebox{.3ex}{$\not\!$}{\vec E}_T$}}
\newcommand{\gevcc}{GeV/$c^2$}
\newcommand{\gevc}{GeV/$c$}
 \def\AuthorList{true} 
 \def\LineNumbers{false}

\begin{document}

\ifthenelse{\equal{\LineNumbers}{true}}{
 \pagewiselinenumbers
 \linenumbers
}

\title{Search for Resonant Top-antitop Production in the Semi-leptonic Decay Mode Using the Full CDF Data Set}

\ifthenelse{\equal{\AuthorList}{true}}{
  \affiliation{Institute of Physics, Academia Sinica, Taipei, Taiwan 11529, Republic of China}
\affiliation{Argonne National Laboratory, Argonne, Illinois 60439, USA}
\affiliation{University of Athens, 157 71 Athens, Greece}
\affiliation{Institut de Fisica d'Altes Energies, ICREA, Universitat Autonoma de Barcelona, E-08193, Bellaterra (Barcelona), Spain}
\affiliation{Baylor University, Waco, Texas 76798, USA}
\affiliation{Istituto Nazionale di Fisica Nucleare Bologna, $^{ee}$University of Bologna, I-40127 Bologna, Italy}
\affiliation{University of California, Davis, Davis, California 95616, USA}
\affiliation{University of California, Los Angeles, Los Angeles, California 90024, USA}
\affiliation{Instituto de Fisica de Cantabria, CSIC-University of Cantabria, 39005 Santander, Spain}
\affiliation{Carnegie Mellon University, Pittsburgh, Pennsylvania 15213, USA}
\affiliation{Enrico Fermi Institute, University of Chicago, Chicago, Illinois 60637, USA}
\affiliation{Comenius University, 842 48 Bratislava, Slovakia; Institute of Experimental Physics, 040 01 Kosice, Slovakia}
\affiliation{Joint Institute for Nuclear Research, RU-141980 Dubna, Russia}
\affiliation{Duke University, Durham, North Carolina 27708, USA}
\affiliation{Fermi National Accelerator Laboratory, Batavia, Illinois 60510, USA}
\affiliation{University of Florida, Gainesville, Florida 32611, USA}
\affiliation{Laboratori Nazionali di Frascati, Istituto Nazionale di Fisica Nucleare, I-00044 Frascati, Italy}
\affiliation{University of Geneva, CH-1211 Geneva 4, Switzerland}
\affiliation{Glasgow University, Glasgow G12 8QQ, United Kingdom}
\affiliation{Harvard University, Cambridge, Massachusetts 02138, USA}
\affiliation{Division of High Energy Physics, Department of Physics, University of Helsinki and Helsinki Institute of Physics, FIN-00014, Helsinki, Finland}
\affiliation{University of Illinois, Urbana, Illinois 61801, USA}
\affiliation{The Johns Hopkins University, Baltimore, Maryland 21218, USA}
\affiliation{Institut f\"{u}r Experimentelle Kernphysik, Karlsruhe Institute of Technology, D-76131 Karlsruhe, Germany}
\affiliation{Center for High Energy Physics: Kyungpook National University, Daegu 702-701, Korea; Seoul National University, Seoul 151-742, Korea; Sungkyunkwan University, Suwon 440-746, Korea; Korea Institute of Science and Technology Information, Daejeon 305-806, Korea; Chonnam National University, Gwangju 500-757, Korea; Chonbuk National University, Jeonju 561-756, Korea; Ewha Womans University, Seoul, 120-750, Korea}
\affiliation{Ernest Orlando Lawrence Berkeley National Laboratory, Berkeley, California 94720, USA}
\affiliation{University of Liverpool, Liverpool L69 7ZE, United Kingdom}
\affiliation{University College London, London WC1E 6BT, United Kingdom}
\affiliation{Centro de Investigaciones Energeticas Medioambientales y Tecnologicas, E-28040 Madrid, Spain}
\affiliation{Massachusetts Institute of Technology, Cambridge, Massachusetts 02139, USA}
\affiliation{Institute of Particle Physics: McGill University, Montr\'{e}al, Qu\'{e}bec H3A~2T8, Canada; Simon Fraser University, Burnaby, British Columbia V5A~1S6, Canada; University of Toronto, Toronto, Ontario M5S~1A7, Canada; and TRIUMF, Vancouver, British Columbia V6T~2A3, Canada}
\affiliation{University of Michigan, Ann Arbor, Michigan 48109, USA}
\affiliation{Michigan State University, East Lansing, Michigan 48824, USA}
\affiliation{Institution for Theoretical and Experimental Physics, ITEP, Moscow 117259, Russia}
\affiliation{University of New Mexico, Albuquerque, New Mexico 87131, USA}
\affiliation{The Ohio State University, Columbus, Ohio 43210, USA}
\affiliation{Okayama University, Okayama 700-8530, Japan}
\affiliation{Osaka City University, Osaka 588, Japan}
\affiliation{University of Oxford, Oxford OX1 3RH, United Kingdom}
\affiliation{Istituto Nazionale di Fisica Nucleare, Sezione di Padova-Trento, $^{ff}$University of Padova, I-35131 Padova, Italy}
\affiliation{University of Pennsylvania, Philadelphia, Pennsylvania 19104, USA}
\affiliation{Istituto Nazionale di Fisica Nucleare Pisa, $^{gg}$University of Pisa, $^{hh}$University of Siena and $^{ii}$Scuola Normale Superiore, I-56127 Pisa, Italy, $^{mm}$INFN Pavia and University of Pavia, I-27100 Pavia, Italy}
\affiliation{University of Pittsburgh, Pittsburgh, Pennsylvania 15260, USA}
\affiliation{Purdue University, West Lafayette, Indiana 47907, USA}
\affiliation{University of Rochester, Rochester, New York 14627, USA}
\affiliation{The Rockefeller University, New York, New York 10065, USA}
\affiliation{Istituto Nazionale di Fisica Nucleare, Sezione di Roma 1, $^{jj}$Sapienza Universit\`{a} di Roma, I-00185 Roma, Italy}
\affiliation{Texas A\&M University, College Station, Texas 77843, USA}
\affiliation{Istituto Nazionale di Fisica Nucleare Trieste/Udine; $^{nn}$University of Trieste, I-34127 Trieste, Italy; $^{kk}$University of Udine, I-33100 Udine, Italy}
\affiliation{University of Tsukuba, Tsukuba, Ibaraki 305, Japan}
\affiliation{Tufts University, Medford, Massachusetts 02155, USA}
\affiliation{University of Virginia, Charlottesville, Virginia 22906, USA}
\affiliation{Waseda University, Tokyo 169, Japan}
\affiliation{Wayne State University, Detroit, Michigan 48201, USA}
\affiliation{University of Wisconsin, Madison, Wisconsin 53706, USA}
\affiliation{Yale University, New Haven, Connecticut 06520, USA}

\author{T.~Aaltonen}
\affiliation{Division of High Energy Physics, Department of Physics, University of Helsinki and Helsinki Institute of Physics, FIN-00014, Helsinki, Finland}
\author{S.~Amerio}
\affiliation{Istituto Nazionale di Fisica Nucleare, Sezione di Padova-Trento, $^{ff}$University of Padova, I-35131 Padova, Italy}
\author{D.~Amidei}
\affiliation{University of Michigan, Ann Arbor, Michigan 48109, USA}
\author{A.~Anastassov$^x$}
\affiliation{Fermi National Accelerator Laboratory, Batavia, Illinois 60510, USA}
\author{A.~Annovi}
\affiliation{Laboratori Nazionali di Frascati, Istituto Nazionale di Fisica Nucleare, I-00044 Frascati, Italy}
\author{J.~Antos}
\affiliation{Comenius University, 842 48 Bratislava, Slovakia; Institute of Experimental Physics, 040 01 Kosice, Slovakia}
\author{G.~Apollinari}
\affiliation{Fermi National Accelerator Laboratory, Batavia, Illinois 60510, USA}
\author{J.A.~Appel}
\affiliation{Fermi National Accelerator Laboratory, Batavia, Illinois 60510, USA}
\author{T.~Arisawa}
\affiliation{Waseda University, Tokyo 169, Japan}
\author{A.~Artikov}
\affiliation{Joint Institute for Nuclear Research, RU-141980 Dubna, Russia}
% ...

\collaboration{CDF Collaboration\footnote{With visitors from
$^a$Istituto Nazionale di Fisica Nucleare, Sezione di Cagliari, 09042 Monserrato (Cagliari), Italy,
$^b$University of California Irvine, Irvine, CA 92697, USA,
$^c$University of California Santa Barbara, Santa Barbara, CA 93106, USA,
$^d$University of California Santa Cruz, Santa Cruz, CA 95064, USA,
$^e$Institute of Physics, Academy of Sciences of the Czech Republic, 182~21, Czech Republic,
$^f$CERN, CH-1211 Geneva, Switzerland,
$^g$Cornell University, Ithaca, NY 14853, USA,
$^h$University of Cyprus, Nicosia CY-1678, Cyprus,
$^i$Office of Science, U.S. Department of Energy, Washington, DC 20585, USA,
$^j$University College Dublin, Dublin 4, Ireland,
$^k$ETH, 8092 Z\"{u}rich, Switzerland,
$^l$University of Fukui, Fukui City, Fukui Prefecture, Japan 910-0017,
$^m$Universidad Iberoamericana, Lomas de Santa Fe, M\'{e}xico, C.P. 01219, Distrito Federal,
$^n$University of Iowa, Iowa City, IA 52242, USA,
$^o$Kinki University, Higashi-Osaka City, Japan 577-8502,
$^p$Kansas State University, Manhattan, KS 66506, USA,
$^q$Brookhaven National Laboratory, Upton, NY 11973, USA,
$^r$University of Manchester, Manchester M13 9PL, United Kingdom,
$^s$Queen Mary, University of London, London, E1 4NS, United Kingdom,
$^t$University of Melbourne, Victoria 3010, Australia,
$^u$Muons, Inc., Batavia, IL 60510, USA,
$^v$Nagasaki Institute of Applied Science, Nagasaki 851-0193, Japan,
$^w$National Research Nuclear University, Moscow 115409, Russia,
$^x$Northwestern University, Evanston, IL 60208, USA,
$^y$University of Notre Dame, Notre Dame, IN 46556, USA,
$^z$Universidad de Oviedo, E-33007 Oviedo, Spain,
$^{aa}$CNRS-IN2P3, Paris, F-75205 France,
$^{bb}$Texas Tech University, Lubbock, TX 79609, USA,
$^{cc}$Universidad Tecnica Federico Santa Maria, 110v Valparaiso, Chile,
$^{dd}$Yarmouk University, Irbid 211-63, Jordan,
$^{ll}$Universite catholique de Louvain, 1348 Louvain-La-Neuve, Belgium,
$^{oo}$University of Z\"{u}rich, 8006 Z\"{u}rich, Switzerland,
$^{pp}$Massachusetts General Hospital and Harvard Medical School, Boston, MA 02114 USA,
$^{qq}$Hampton University, Hampton, VA 23668, USA,
$^{rr}$Los Alamos National Laboratory, Los Alamos, NM 87544, USA
}}
\noaffiliation
 
}

\ifthenelse{\equal{\AuthorList}{false}}{
  \begin{figure}  
  \leftline{\includegraphics[scale=0.5]{cdfii_thumb_logo}\hfill CDF/PUB/EXOTIC/CDFR/10906}
  \end{figure}
\vspace*{0.5cm} 
\author{The CDF Collaboration}
\affiliation{URL http://www-cdf.fnal.gov}
}

\date{\today}

\begin{abstract}
  This Letter reports a search for a narrow resonant state decaying
into two $W$ bosons and a bottom-antibottom quark pair where one $W$ boson decays leptonically and the other
decays into a quark-antiquark pair.   The search is
particularly sensitive to top-antitop resonant production.  We use the full data
sample of proton-antiproton collisions at a center-of-mass energy of
1.96 TeV collected by the CDF II detector at the Fermilab Tevatron,
corresponding to an integrated luminosity of 9.45 fb$^{-1}$. No
evidence for resonant production is found and upper limits
on the production cross section times branching ratio for a narrow
resonant state are extracted.  Within a specific benchmark model, we exclude a 
$Z'$ boson with mass below 915~\gevcc\ decaying into a top-antitop pair
at the 95\% credibility level assuming a $Z'$ boson decay width of
$\Gamma_{Z'} = 0.012~M_{Z'}$. This is the most sensitive search for a 
narrow $\ensuremath{q\bar{q}}$-initiated $\ensuremath{t\bar{t}}$ resonance 
in the mass region below 750~\gevcc.

\end{abstract}
\pacs{13.85.Rm, 14.65.Ha, 14.70.Pw, 14.80.Tt}

\maketitle

%==================================================================

   The large mass of the top quark, compared to that of the other
fundamental particles, gives it a special position within the standard
model (SM). Since its discovery~\cite{TopDiscovery}, the top quark has
played an important role in theoretical extensions beyond the standard
model~(BSM)~\cite{Hill:1993hs}.  Recently, renewed interest has been
directed toward searches including top quark final states for BSM physics due to discrepancies reported in the
$\ttb$ forward-backward
asymmetry~\cite{Aaltonen:2011kc,Kamenik:2011wt}. Moreover, the most
recent search for resonant $\ttb$ production from
D0~\cite{Abazov:2011gv} reports an approximately 2$\sigma$ excess of events at resonant-mass values around 950~\gevcc.  Many BSM theories~\cite{Frederix,Dobrescu:2009vz,Gresham:2011pa,Agashe:2006hk,Bai:2011ed}
predict heavy resonances that add a resonant component to the SM
$\ttb$ production mechanism.

  Top quarks decay via the weak interaction, nearly always into a $W$
boson and a $b$ quark. The $W$ bosons then decay into lighter
fermion-antifermion pairs~\cite{pdgstats}.  A leptonic decay into a
charged lepton and a neutrino occurs 32\% of the time while a
hadronic decay into an up-type quark and a down-type quark occurs the
remaining 68\% of the time.  We search for resonant production of
top quark pairs followed by decays into a final state with one lepton
and multiple jets, where one of the $W$ bosons decays leptonically (to
either an electron or a muon plus a neutrino) and the other $W$ boson decays hadronically.
This semi-leptonic channel features a distinctive final state due to
the presence of a charged lepton and has a branching ratio of 29\%.
  
  Unlike previous searches at CDF~\cite{CDFRunI,ZprimeCDF,Aaltonen:2011vi,Aaltonen:2011ts}, we do not apply constraints based on the presence of top quarks in the event. While we focus the discussion on \ttb\ resonances, we construct the top-antitop mass \mttb\ used as a final search discriminant by taking the invariant mass of all objects~(lepton, jets, and  missing $E_T$) in the event including those that may not originate from top quark production. Other than the event selection defined below, which provides a sample primarily composed of \ttb\ events, there are no requirements that the event be consistent with \ttb\ production.  This results in a more general search that is sensitive not only to \ttb, but also to any heavy narrow resonance decaying into a final state with a $W$ boson and three or more jets with one or two jets originating from a $b$ quark.

  As a benchmark model, we consider a specific SM extension,
topcolor-assisted technicolor~\cite{Hill2}. This model explains the
large mass of the top quark through the introduction of new strong
dynamics and also predicts a vector particle ($Z'$ boson), which
couples primarily to the third generation of quarks and has no
significant couplings to leptons. The existence of a narrow-width 
$Z'$ boson resonance~($\Gamma_{Z'} = 0.012~M_{Z'}$) decaying to $\ttb$
pairs, using the leptophobic topcolor model~\cite{Harris}, has been
searched for both by the
CDF~\cite{CDFRunI,ZprimeCDF,Aaltonen:2011vi,Aaltonen:2011ts} and
D0~\cite{D0RunI,ZprimeD0,Abazov:2011gv} experiments at the Tevatron,
and also by the ATLAS~\cite{Aad:2012tx} and
CMS~\cite{CMS:2012,CMS:2012cx} experiments at the LHC.  For resonance
searches at the highest masses, the LHC experiments have
superior sensitivity to the Tevatron due to the higher center-of-mass
energy.  However, in the lower mass regions ($m_{Z'} <$~750~\gevcc)
the Tevatron experiments have competitive sensitivity in
searches for particles produced in \qqb-initiated states, such as the
$Z'$ boson.  While the production rate for the main background from SM
\ttb\ production is approximately 25 times larger, no
valence antiquarks are available in the LHC $pp$ collisions, so the
signal production rate increases by a smaller factor relative to the
\ppb\ collisions of the Tevatron (between four and eight depending on
the signal mass hypothesis).

 The collision events discussed in this Letter were produced at
 the Tevatron \ppb\ collider at a center-of-mass energy of 1.96 TeV and were
 recorded by the CDF II detector~\cite{CDFII}. The data sample
 corresponds to the full data set of the Tevatron, which comes from an
 integrated luminosity of 9.45 fb$^{-1}$. The CDF II detector consists
 of high-precision tracking systems for vertex and charged-particle
 track reconstruction, surrounded by electromagnetic and hadronic
 calorimeters for energy measurement, and muon subsystems outside the
 calorimeter for muon detection. CDF II uses a cylindrical coordinate
 system with azimuthal angle $\phi$, polar angle $\theta$ measured
 with respect to the positive $z$ direction along the proton beam, and
 the distance $r$ measured from the beamline. The pseudorapidity, transverse
 energy, and transverse momentum are defined as
 $\eta=-\ln\left[\tan(\frac{\theta}{2})\right]$,
 $E_{T}=E\sin{\theta}$, and $p_{T}=p\sin{\theta}$, respectively, where
 $E$ and $p$ are the energy and momentum of an outgoing particle.  The
 missing transverse energy $\METVEC$ is defined by $\METVEC =-\sum_iE_T^i{\hat
 n}_i$, where ${\hat n}_i$ is a unit vector perpendicular to the beam
 axis that points to the $i$th calorimeter tower ($\MET=|\METVEC|$). 

 The event selection and background estimation methods summarized
 below closely follow those that were employed in the observation of single top quark
 production~\cite{Aaltonen:2010jr} and in the search for the Higgs boson
 in the $WH \rightarrow \ell\nu b\bar{b}$ final
 state~\cite{Aaltonen:2012tf}.  The main difference with respect to
 the current search is the jet multiplicity requirement.

The data were collected using online event selections (triggers)
  requiring one of the following energetic-lepton signatures: a
  high transverse momentum ($p_T$) electron candidate, a high-$p_T$ muon candidate, or large
  $\MET$.  Significant $\MET$ can be produced when the neutrino from a
  leptonic $W$ boson decay escapes detection.

Candidate events are selected by requiring a lepton candidate with
$p_T^{\ell}>20$~\gevc,~~$\MET>20$~GeV, and three or more jets with
$|\eta|<2.0$ and $E_T>20$~GeV after correcting the jet energies for
instrumental effects~\cite{JetClu,nimjes}.  One or two jets
must be identified as being likely to have originated from a $b$ quark
according to the {\sc secvtx}~\cite{secvtx} algorithm. This algorithm
searches in the jet for a secondary vertex which  results from the
displaced decay of $B$ hadrons.  Events are rejected if more than one identified
lepton is reconstructed, or if they are kinematically inconsistent with
leptonic $W$ boson decays~\cite{Sforza:2011zz}.  Events with severely
misreconstructed jets or leptons are removed based on angular
correlations between the jet or lepton candidate and the \METVEC.

  Models for background processes are derived from a mixture of
simulation and data-driven techniques~\cite{Aaltonen:2010jr}.
Important backgrounds in this final state include SM \ttb\ production
and other processes that include a $W$ or $Z$ boson in association with three
or more jets. The events can include true $b$-quark jets, as in $W$
boson + $b\bar{b}j$ events, or jets that have been misidentified as
$b$-quark jets, such as in $W$ boson $ + c\bar{c}j$ and $W$ boson $+
jjj$ events, where $j$ refers to jets not originating from
heavy-flavor quarks.  Multijet events without $W$ bosons also
contribute to the sample composition.  Additional small background contributions come from $Z$ boson production with
additional jets, diboson production, and single top quark production.

The expected rate for the SM \ttb\ background is taken to be $7.04\pm
0.50$~pb~\cite{mochuwer} as calculated at next-to-next-to-leading
order using MSTW 2008 parton distribution functions~\cite{MSTW}. In
order to predict the acceptance for non-resonant SM \ttb\ events 
and their kinematic distributions, we use a sample of Monte Carlo~(MC)
events generated using {\sc powheg}~\cite{powheg} and assuming a
top quark mass of 172.5~\gevcc~\cite{topmass} with parton showering
provided by {\sc pythia} v6.2~\cite{pythia} followed by simulation of
the CDF II detector~\cite{GEANT3,GFLASH}. The detection efficiency
predicted by the MC is corrected based on measurements using data for
lepton identification, trigger efficiencies, and $b$-jet tagging
efficiencies.  The normalization for the $W$ boson + jet processes is
obtained from a fit to the \MET\ distribution before the $b$-tagging
requirement is applied.  The background from events with mistakenly
$b$-tagged light-flavor jets, $W$ boson $+ jjj$ for example, is
estimated by measuring the rate of such mistags in multijet
data~\cite{secvtx} and applying this rate to the $W$ boson + jets data
samples before $b$ tagging.  The contribution from true heavy-flavor
production in the $W$ boson + jets event sample is determined from
measurements of the heavy-flavor event fraction in a $W$ boson + 1 jet
sample that is independent of the sample used in the resonance search.
We model the kinematic distributions of $W$ boson + jets events using a combination
of {\sc{alpgen}}~\cite{Mangano:2002ea} matrix-element generation and
{\sc{pythia}} parton showering. The rate of the QCD multijet
background is obtained from a fit to the \MET\ distribution, where the
QCD multijet background is modeled using a sample of collision events
in which one of the lepton identification requirements is inverted to
obtain an enriched sample of QCD multijet events.

  The background predictions are summarized in
Table~\ref{table:MethodII}.  In this table and the following figures
we have divided the sample into events that include three jets and
events that include four or more jets.  For the statistical
interpretation of the data we further subdivide the events based on the
number of $b$-tagged jets (one or two $b$ tags) and based on the
lepton type (lepton types that can be directly identified by the
trigger, or leptons in events selected with the
\MET-based trigger), yielding eight independent channels used to
search for a resonance in the \mttb\ distributions.  The sensitivity
of the search benefits from this subdivision because the search
subchannels have different background compositions,
signal-to-background ratios, and invariant mass resolutions.

% Start 3JET tag=SVLSVL leptype=CEMCMUPCMXISOTRK for background table 
\begin{table}
\begin{center}
\begin{ruledtabular}
\begin{tabular}{cccc}
Process                     &   3-jet events        &   $\ge4$-jet events      \\
\hline						
\ttb							&	1925	$\pm$	204     &	2565	$\pm$	271  	\\
$W/Z$ boson + jets					&	2281	$\pm$	607 	&	569	$\pm$	189  	\\
Multijets						&	147	$\pm$	60  	&	126	$\pm$	104  	\\					
\hline
Total background&	4354	$\pm$	872 	&	3260	$\pm$	563  	\\
\hline
Observed					&	4254				  	&	3049			         \\
\end{tabular}
\end{ruledtabular}
\caption{Summary of the background prediction and observed data for three-jet and four-or-more-jet events. The uncertainties include statistical and systematic contributions.}
\label{table:MethodII}
\end{center}
\end{table}
% End 3JET tag=SVLSVL leptype=CEMCMUPCMXISOTRK for background table 

We use the invariant mass of all reconstructed objects in the event to discriminate between SM background and $Z'$ boson
signal events. For each event we calculate \mttb\ using the momenta
of the three or more jets, the charged lepton, and the
neutrino. The transverse momentum of the neutrino is estimated from
the $\METVEC$.  However, because the $z$-component of the momenta of
the scattering partons from the \ppb\ collision is unknown, the
final-state reconstructed energy need not be balanced in
the $z$ direction.  The longitudinal component of the neutrino
momentum ($p_z$) is determined by solving $M_{W}^2 = (p^l +
p^{\nu})^2$, where $M_{W}$, $p^l$, and $p^{\nu}$ are the $W$ boson mass, the lepton momentum, and the neutrino momentum, respectively. 
The smaller solution of the resulting quadratic equation is chosen for the $p_z$ of the neutrino.  If there is no real
solution we set neutrino $p_z$ = 0.  This approach is found to select
the correct $p_z$ of the neutrino in about 70\% of simulated \ttb\ events.

\begin{table}[htbp]

\begin{center}
\begin{ruledtabular}
\begin{tabular}{c c c c}
                     &     \multicolumn{2}{c}{Acceptance} &  $\sigma_{Z'} BR(Z'\to t\bar{t})$ \\
$M_{Z'}$ [\gevcc] & 3 jets [\%] & $\geq$4 jets [\%] &  [pb]\\ 
\hline
350&2.75 $\pm $ 0.15&3.68 $\pm $ 0.21&8.91\\
400&2.82 $\pm $ 0.16&3.83 $\pm $ 0.21&12.3\\
450&2.35 $\pm $ 0.13&3.47 $\pm $ 0.19&8.24\\
500&2.29 $\pm $ 0.13&3.59 $\pm $ 0.20&5.53\\
550&2.16 $\pm $ 0.12&3.63 $\pm $ 0.21&3.51\\
600&1.93 $\pm $ 0.11&3.58 $\pm $ 0.21&2.30\\
650&1.71 $\pm $ 0.10&3.48 $\pm $ 0.20&1.43\\
700&1.52 $\pm $ 0.09&3.24 $\pm $ 0.19&0.917\\
750&1.37 $\pm $ 0.09&2.99 $\pm $ 0.18&0.566\\
800&1.19 $\pm $ 0.08&2.75 $\pm $ 0.17&0.355\\
850&1.10 $\pm $ 0.07&2.46 $\pm $ 0.15&0.208\\
900&0.96 $\pm $ 0.06&2.24 $\pm $ 0.14&0.134\\
950&0.92 $\pm $ 0.06&2.03 $\pm $ 0.12&0.080\\
1000&0.87 $\pm $ 0.06&1.91 $\pm $ 0.12&0.049\\
1100&0.86 $\pm $ 0.06&1.84 $\pm $ 0.11&0.017\\
1200&1.05 $\pm $ 0.07&2.01 $\pm $ 0.12&0.006\\
%\hline
\end{tabular}
\end{ruledtabular}
\caption{Acceptances and predicted cross sections times the branching ratio for the leptophobic topcolor $Z'$ boson signal hypotheses.}\label{tab:accept}
\end{center}
\end{table}

\begin{figure*}[ht]
  \begin{center}
  \includegraphics[width=3.2in]{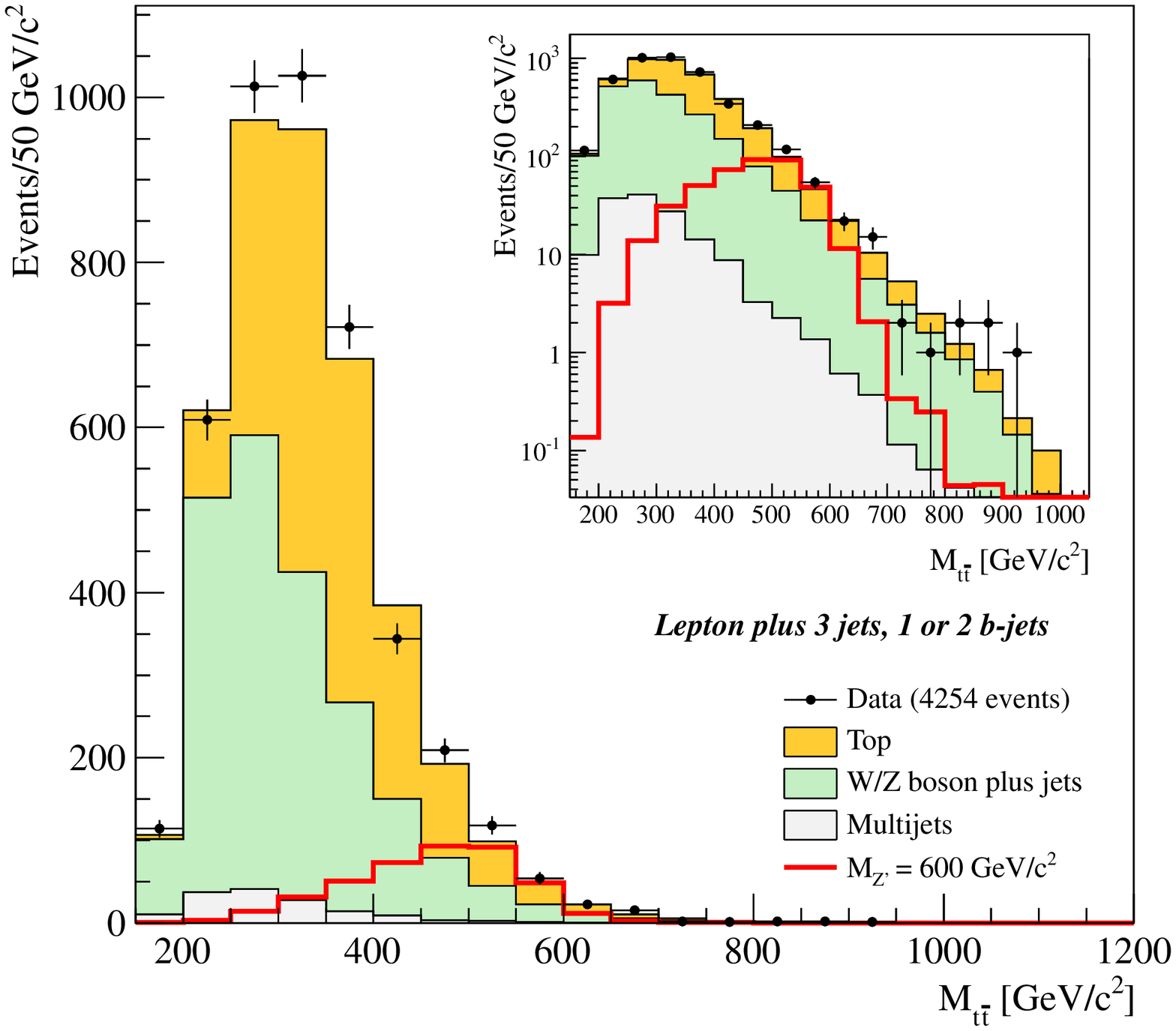}
  \includegraphics[width=3.2in]{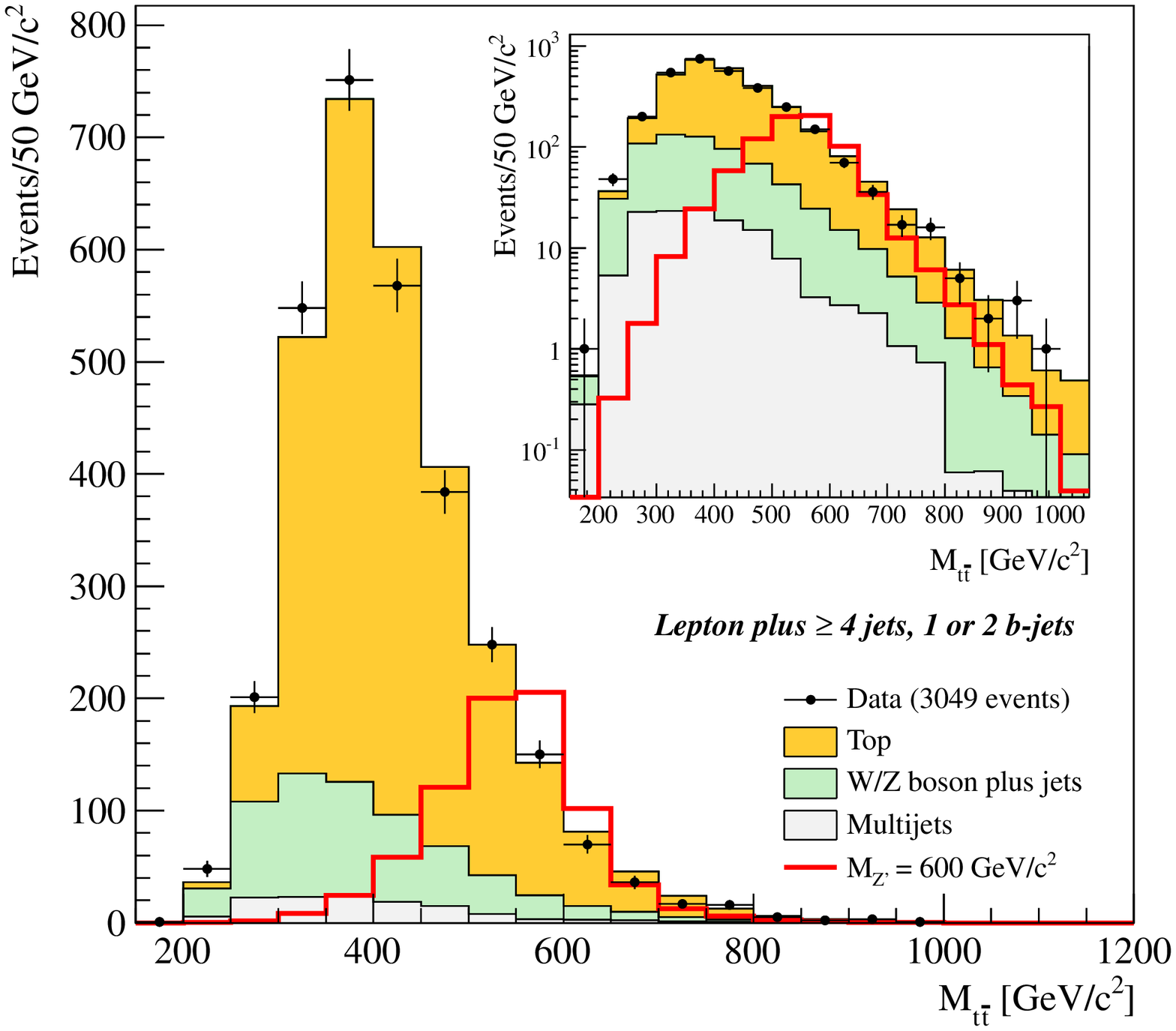}
     \caption{Reconstructed $\mttb$ for the three-jet events (left) and four-or-more-jet events (right). The full distribution is on a linear scale while the high-mass tail in the inset is drawn on a logarithmic scale.  The background expectation is normalized to the best fit from the data. The red histogram shows the expectation for a 600~\gevcc\ leptophobic topcolor resonance normalized to the predicted cross section.}
    \label{fig:Data}
  \end{center}
\end{figure*}

For the benchmark model, the $Z'$ boson cross sections times branching
 fraction are based on leading-order predictions from
 Ref.~\cite{MttXsec} with an additional scaling factor of 1.3 applied to
 account for next-to-leading-order (NLO)
 effects~\cite{Gao:2010bb}. Signal $Z'$ boson events are modeled with
 simulated events generated by {\sc pythia} in order to study the
 signal acceptance and to predict the \mttb\ distributions.
 Table~\ref{tab:accept} shows the selection efficiencies and cross
 sections for $Z'$ boson events after the final event selection for
 each mass hypothesis considered in the analysis.

A total of 4254 (3049) events survive the selection criteria
for the three-jet (four-or-more-jet) category.  The SM
$\ttb$ contribution is estimated to be 43\% (78\%) for three-jet
(four-or-more-jet) events. The remaining events are contributed primarily from the
$W$ boson + jet and QCD multijet processes plus a potential signal
contribution from $Z'$ boson events.  The \mttb\ distributions for the
background model and events observed in the data are shown in Fig.~\ref{fig:Data}.  The
 \mttb\ distribution for the $Z'$ boson signal for the 600~\gevcc\ mass
 hypothesis is also included in Fig.~\ref{fig:Data}.

We calculate a Bayesian credibility level (C.L.) limit on resonant
$\ttb$ production for each mass hypothesis based on the binned observed
$\mttb$ spectrum using the combined likelihood which includes the priors, $\pi({\vec{\theta}}$, on the systematic uncertainties, $\vec{\theta}$: 

\begin{equation}
{\cal{L}}(R,{\vec{s}},{\vec{b}}|{\vec{n}},{\vec{\theta}})\times\pi({\vec{\theta}})
= \prod_{i=1}^{N_{\rm C}}\prod_{j=1}^{N_{\rm bins}} \mu_{ij}^{n_{ij}} \frac{e^{-\mu_{ij}}}{n_{ij}!}
\times\prod_{k=1}^{n_{\rm sys}}e^{-\theta_k^2/2}.
\end{equation}
\noindent In this expression, the first product is over the number of channels
$N_{\rm C}$, and the second product is over histogram bins
containing $n_{ij}$ events.  The predictions for the
bin contents are $\mu_{ij} =R \times s_{ij}({\vec{\theta}}) +
b_{ij}({\vec{\theta}})$ for channel $i$ and histogram bin $j$, where
$s_{ij}$ represents the potential resonant signal, $b_{ij}$
is the expected background in the bin, and $R$ is a scaling
factor applied to the signal.

Systematic uncertainties are parametrized by the dependence of
$s_{ij}$ and $b_{ij}$ on ${\vec\theta}$.  Each of the $n_{\rm sys}$
components of ${\vec\theta}$, $\theta_k$, corresponds to a single
independent source of systematic uncertainty. We account for correlations by allowing each parameter to have an impact on several sources of signal and background in
different channels.  Gaussian priors
are assumed for the $\theta_k$, which are truncated so that no prediction is
negative.  The likelihood function, multiplied by the $\theta_k$
priors, $\pi(\theta_k)$, is then integrated over $\theta_k$ including
correlations~\cite{pdgstats}:

\begin{equation}
{\cal{L}}^\prime(R) = \int{\cal{L}}(R,{\vec{s}},{\vec{b}}|{\vec{n}},{\vec{\theta}})\pi({\vec{\theta}})d{\vec{\theta}}.
\end{equation}

We assume a uniform prior in $R$ to obtain its posterior distribution.
The observed 95\% C.L. upper limit on $R$, $R_{95}^{\rm{obs}}$,
satisfies $0.95=\int_0^{R_{95}^{\rm{obs}}}\cal{L}^\prime(R)
\mbox{d}R$.  The expected distribution of $R_{95}$ is computed in an
ensemble of pseudoexperiments generated without signal.  In each
pseudoexperiment, values of the nuisance parameters are drawn
from their priors.  The median expected value of $R_{95}$ in this
ensemble is quoted as the expected limit.  This statistical procedure
is repeated for each resonance-mass hypothesis from 350~\gevcc\ to
1200~\gevcc.

We consider uncertainties that affect the normalization as well as
 uncertainties that affect the \mttb\ distributions.  The same set of
 uncertainties on the dominant background (SM \ttb\ production) and
 the resonant signal are considered: they arise from the
 uncertainty in the jet energy scale (JES)~\cite{nimjes}, the
 $b$-tagging efficiency, the luminosity measurement~\cite{lumi}, the
 lepton identification and trigger efficiency (2--6\%), and the rate of
 initial- and final-state (IFSR) radiation from the parton shower
 model.  The JES, $b$-tag, and IFSR variations also affect the shape
 of the \mttb\ distributions.  The rate of production for events with
 a $W$ boson and heavy-flavor jet ($b$ or $c$) is assigned an uncertainty of 30\%
 due to limitations in the calibration of the fraction of
 heavy-flavor jets in the sample.  Uncertainties on the
 renormalization and factorization scale used in the {\sc alpgen}
 sample affect the shape of the \mttb\ distributions from $W$ boson +
 jets. The QCD multijet background normalization is assigned a 40\% 
uncertainty due to statistical limitations from the fitting procedure
 and the definition of the multijet model~\cite{Aaltonen:2010jr}.

\begin{figure}[htbp]
  \begin{center}
    \includegraphics[width=3.5in]{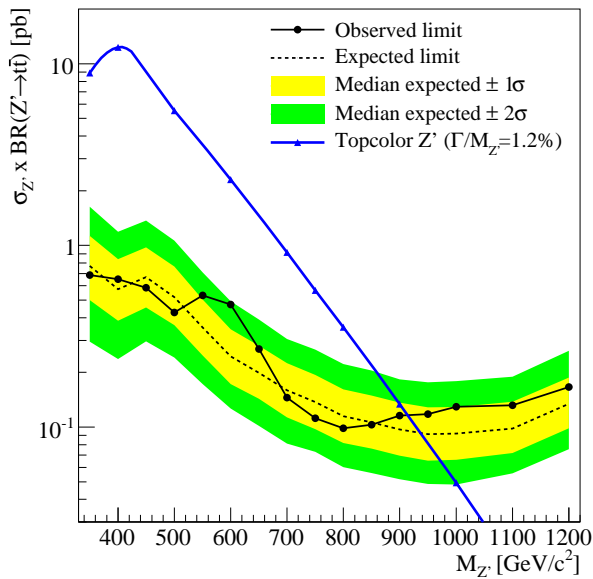}
    \caption{
      Expected and observed upper limits on the production cross section times the branching 
      ratio for the leptophobic topcolor $Z'$ boson in 9.45 fb$^{-1}$ of CDF data. The dashed line
      is the median expected upper limit with the assumption of no signal, the black points
      are the observed limit, and the blue line is the cross section prediction for leptophobic
      topcolor $Z'$ boson production.}
    \label{fig:Brazil}
  \end{center}
\end{figure}

The resulting 95\% C.L. upper limits on $\sigma(p\bar p \rightarrow
Z') BR(Z' \rightarrow t\bar t )$ as a function of $\mttb$ are
shown in Fig.~\ref{fig:Brazil} and Table~\ref{table:Limit} together
with expected limits derived from pseudoexperiments that include the
SM background hypothesis only.  A benchmark leptophobic topcolor model
is excluded at 95\% C.L. for $Z'$ boson masses smaller than 915~\gevcc\, 
assuming the width of the resonance is $\Gamma_{Z'} =
0.012~M_{Z'}$.  In addition, the limits reported here can be applied to
any resonance producing the same final state as long as the decay width is significantly smaller than the reconstruction mass resolution ($\Gamma_{Z'} \ll 0.15~M_{Z'}$), and the
difference in the acceptance with respect to the values quoted in
Table~\ref{tab:accept} is taken into account.

\begin{table}
\begin{center}
\begin{ruledtabular}
\begin{tabular}{ccc}
$M_{Z'}$ [\gevcc]& Expected [pb]& Observed [pb] \\
\hline
350 &0.772 & 0.687 \\
400 &0.575 & 0.652 \\
450 &0.670 & 0.585 \\
500 &0.520 & 0.427 \\
550 &0.354 & 0.530 \\
600 &0.245 & 0.472 \\
650 &0.199 & 0.269 \\
700 &0.159 & 0.145 \\
750 &0.137 & 0.112 \\
800 &0.115 & 0.099 \\
850 &0.106 & 0.103 \\
900 &0.097 & 0.116 \\
950 &0.091 & 0.118 \\
1000 &0.092 & 0.129 \\
1100 &0.098 & 0.132 \\
1200 &0.134 & 0.166 \\
\end{tabular}
\end{ruledtabular}
\caption{Expected and observed 95\% C.L. upper limits on the production cross section times the branching ratio for a narrow \ttb\ resonance, given as a function of $Z'$ boson mass.}
\label{table:Limit}
\end{center}
\end{table}

In conclusion, we have performed a search for a heavy resonance
decaying into $\ttb$ using  the semi-leptonic decay channel in data from 9.45 fb$^{-1}$ 
of integrated luminosity.  The data are found to be consistent with the background 
expectation and upper limits are set on the production cross section times 
branching ratio at the 95\% C.L. For a specific benchmark model (leptophobic topcolor), 
we exclude $Z'$ bosons with masses up to 915~\gevcc. For masses smaller than approximately 
$750$~\gevcc, this search yields the most constraining limits 
to date on \qqb-produced narrow \ttb\ resonant states in the semi-leptonic decay mode.

\begin{center}
{\bf Acknowledgments}
\end{center}

We thank the Fermilab staff and the technical staffs of the
participating institutions for their vital contributions. This work
was supported by the U.S. Department of Energy and National Science
Foundation; the Italian Istituto Nazionale di Fisica Nucleare; the
Ministry of Education, Culture, Sports, Science and Technology of
Japan; the Natural Sciences and Engineering Research Council of
Canada; the National Science Council of the Republic of China; the
Swiss National Science Foundation; the A.P. Sloan Foundation; the
Bundesministerium f\"ur Bildung und Forschung, Germany; the Korean
World Class University Program, the National Research Foundation of
Korea; the Science and Technology Facilities Council and the Royal
Society, UK; the Russian Foundation for Basic Research; the Ministerio
de Ciencia e Innovaci\'{o}n, and Programa Consolider-Ingenio 2010,
Spain; the Slovak R\&D Agency; the Academy of Finland; and the
Australian Research Council (ARC).

\end{document}